\newfam\scrfam
\batchmode\font\tenscr=rsfs10 \errorstopmode
\ifx\tenscr\nullfont
	\message{rsfs script font not available. Replacing with calligraphic.}
\else	\font\sevenscr=rsfs7 
	\font\fivescr=rsfs5 
	\skewchar\tenscr='177 \skewchar\sevenscr='177 \skewchar\fivescr='177
	\textfont\scrfam=\tenscr \scriptfont\scrfam=\sevenscr
	\scriptscriptfont\scrfam=\fivescr
	\def\scr{\fam\scrfam}
	\def\cal{\scr}
\fi
\newfam\msbfam
\batchmode\font\twelvemsb=msbm10 scaled\magstep1 \errorstopmode
\ifx\twelvemsb\nullfont\def\Bbb{\bf}
	\font\fourteenbbb=cmb10 at 14pt
	\font\eightbbb=cmb10 at 8pt
	\message{Blackboard bold not available. Replacing with boldface.}
\else	\catcode`\@=11
	\font\tenmsb=msbm10 \font\sevenmsb=msbm7 \font\fivemsb=msbm5
	\textfont\msbfam=\tenmsb
	\scriptfont\msbfam=\sevenmsb \scriptscriptfont\msbfam=\fivemsb
	\def\Bbb{\relax\expandafter\Bbb@}
	\def\Bbb@#1{{\Bbb@@{#1}}}
	\def\Bbb@@#1{\fam\msbfam\relax#1}
	\catcode`\@=\active
	\font\fourteenbbb=msbm10 at 14pt
	\font\eightbbb=msbm8
\fi

\font\eightrm=cmr8		\def\xrm{\eightrm}
\font\eightbf=cmbx8		\def\xbf{\eightbf}
\font\eightit=cmti8	
				\def\xit{\eightit}
\font\eighttt=cmtt8		\def\xtt{\eighttt}
\font\eightcp=cmcsc8
\font\eighti=cmmi8		\def\xold{\eighti}
\font\teni=cmmi10		\def\old{\teni}

\font\twelverm=cmr12

\font\twelvecp=cmcsc10 scaled\magstep1
\font\fourteencp=cmcsc10 scaled\magstep2
\font\fiverm=cmr5		\def\trm{\fiverm}
\font\twelvemath=cmmi12

\font\fourteenrm=cmr12 at 14pt

\font\eightmath=cmmi8
\def\ss{\scriptstyle}

\def\bZ{{\Bbb Z}}
\headline={\ifnum\pageno=1\hfill\else
{\eightcp M. Cederwall and A. Westerberg: 
``World-volume fields, 
{\eightrm SL}(2;{\eightbbb Z}) 
		and duality$\ldots$''}\dotfill{ }{\old\folio}\fi}
\def\makeheadline{\vbox to 0pt{\vss\noindent\the\headline\break
\hbox to\hsize{\hfill}}
	\vskip1.65\baselineskip}
\def\makefootline{\ifnum\foottest=1
	\baselineskip=.8cm\line{\the\footline}\global\foottest=0
	\fi
        }
\newcount\foottest
\global\foottest=0

\newcount\refcount
\refcount=0
\newwrite\refwrite
\def\ref#1#2{\global\advance\refcount by 1
	\xdef#1{{\old\the\refcount}}
	\ifnum\the\refcount=1
	\immediate\openout\refwrite=\jobname.refs
	\fi
	\immediate\write\refwrite
		{\item{[{\xold\the\refcount}]} #2\hfill\par\vskip-2pt}}
\def\refout{\catcode`\@=11 
	\xrm\immediate\closeout\refwrite
	\vskip1.5\baselineskip
	{\noindent\twelvecp References}\hfill\vskip.5\baselineskip
	\parskip=.875\parskip 
	\baselineskip=.8\baselineskip
	\input\jobname.refs 
	\parskip=8\parskip \divide\parskip by 7
	\baselineskip=1.25\baselineskip 
	\catcode`\@=\active\rm}
\newcount\sectioncount
\sectioncount=0
\def\section#1#2{\global\eqcount=0
	\global\advance\sectioncount by 1
	\vskip\baselineskip\noindent
	\hbox{\twelvecp\the\sectioncount. #2\hfill}\vskip.3\baselineskip
	\xdef#1{{\old\the\sectioncount}}\noindent}
\newcount\appendixcount
\appendixcount=0
\def\appendix#1{\global\aeqcount=0
	\global\advance\appendixcount by 1
	\vskip\baselineskip\noindent
	\ifnum\the\appendixcount=1
	\hbox{\twelvecp Appendix A. #1\hfill}\vskip.3\baselineskip\fi
    \ifnum\the\appendixcount=2
	\hbox{\twelvecp Appendix B. #1\hfill}\vskip.3\baselineskip\fi
    \ifnum\the\appendixcount=3
	\hbox{\twelvecp Appendix C. #1\hfill}\vskip.3\baselineskip\fi\noindent}
\newcount\eqcount
\eqcount=0
\def\Eqn#1{\global\advance\eqcount by 1
	\xdef#1{{\old\the\sectioncount}.{\old\the\eqcount}}
		\eqno({\oldstyle\the\sectioncount}.
		{\oldstyle\the\eqcount})}
\def\eqn{\global\advance\eqcount by 1
	\eqno({\oldstyle\the\sectioncount}.{\oldstyle\the\eqcount})}
\def\multi{\global\advance\eqcount by 1}
\def\multieq#1#2{\xdef#1{{\old\the\eqcount#2}}
	\eqno{({\oldstyle\the\eqcount#2})}}
\newcount\aeqcount
\aeqcount=0
\def\AEqn#1{\global\advance\aeqcount by 1
	\ifnum\the\appendixcount=1 
		\xdef#1{\hbox{\xrm A}.{\old\the\aeqcount}}
		\eqno(\hbox{\xrm A}.{\oldstyle\the\aeqcount})\fi
	\ifnum\the\appendixcount=2
		\xdef#1{\hbox{\xrm B}.{\old\the\aeqcount}}
		\eqno(\hbox{\xrm B}.{\oldstyle\the\aeqcount})\fi
	\ifnum\the\appendixcount=3
		\xdef#1{\hbox{\xrm C}.{\old\the\aeqcount}}
		\eqno(\hbox{\xrm C}.{\oldstyle\the\aeqcount})\fi}
\def\aeqn{\global\advance\aeqcount by 1
	\ifnum\the\appendixcount=1 
		\eqno(\hbox{\xrm A}.{\oldstyle\the\aeqcount})\fi
	\ifnum\the\appendixcount=2
		\eqno(\hbox{\xrm B}.{\oldstyle\the\aeqcount})\fi
	\ifnum\the\appendixcount=3
		\eqno(\hbox{\xrm C}.{\oldstyle\the\aeqcount})\fi}
\parskip=3.5pt plus .3pt minus .3pt
\baselineskip=14pt plus .5pt minus .1pt
\lineskip=.5pt plus .05pt minus .05pt
\lineskiplimit=.5pt
\abovedisplayskip=19pt plus 4pt minus 2pt
\belowdisplayskip=\abovedisplayskip
\hsize=15cm
\vsize=20.5cm
\hoffset=1cm
\voffset.5cm

\def\noblackbox{\overfullrule=0pt}
\noblackbox

\frenchspacing

\def\/{\over}
\def\*{\partial}
\def\a{\alpha}
\def\b{\beta}

\def\d{\delta}
\def\e{\varepsilon}

\def\g{\gamma}
\def\k{\kappa}
\def\l{\lambda}

\def\F{{\cal F}}

\def\G{\Gamma}
\def\L{{\cal L}}

\def\R{{\Bbb R}}

\def\Z{{\Bbb Z}}
\def\C{{\Bbb C}}

\def\punkt{\,.}
\def\komma{\,,}
\def\semikolon{\,;}
\def\is{\!=\!}
\def\minus{\!-\!}
\def\plus{\!+\!}
\def\>{\!>\!}
\def\w{\!\wedge\!}
\def\dot{\!\cdot\!}

\def\half{{\lower2.5pt\hbox{\eightrm 1}\/\raise2.5pt\hbox{\eightrm 2}}}
\def\frac#1{{\lower2.5pt\hbox{\eightrm1}\/\raise2.5pt\hbox{\eightrm#1}}}
\def\tfrac#1{{\lower2.5pt\hbox{\trm1}\/\raise2.5pt\hbox{\trm#1}}}
\def\genfrac#1#2{{\lower2.5pt\hbox{\eightrm#1}\/\raise2.5pt\hbox{\eightrm#2}}}

\def\tr{\hbox{\rm tr}}
\def\ie{{i.e.,}}
\def\eg{e.g.}

\def\cc{\hbox{c.c.}}

\def\det{\hbox{\rm det}}
\def\II{\hbox{I\hskip-0.6pt I}}

\def\id{1\!\!1}

\def\C{{\cal C}}
\def\A{{\cal A}}
\def\F{{\cal F}}
\def\G{{\cal G}}
\def\H{{\cal H}}
\def\U{{\cal U}}
\def\K{{\cal K}}

\def\Im{\hbox{Im}\,}
\def\Re{\hbox{Re}\,}

\def\ihalf{{\lower2.5pt\hbox{\eightmath i}\/\raise2.5pt\hbox{\eightrm 2}}}
\def\ifrac#1{{\lower2.5pt\hbox{\eightmath i}\/\raise2.5pt\hbox{\eightrm#1}}}

\def\ccr{\cr\noalign{\smallskip}}
\def\cccr{\cr\noalign{\smallskip\smallskip}}

%
%
%
%

\null\vskip-1.5cm

\hfill\vbox{\hbox{G\"oteborg-ITP-97-14}
\hbox{DAMTP-97-107}
\hbox{\tt 
hep-th/9710007}
\hbox{October 1, 1997}}

\vskip2.5cm
\centerline{\fourteencp World-volume fields, 
{\fourteenrm SL(2;}{\fourteenbbb Z}{\fourteenrm )} 
	and duality:} 
\vskip4pt
\centerline{\fourteencp The type \II B 3-brane}
\vskip\parskip
\centerline{\twelvecp}

\vskip1.2cm
\centerline{\twelverm Martin Cederwall} 
\vskip.6cm
\catcode`\@=11 
\centerline{\it Institute for Theoretical Physics,}
\centerline{\it G\"oteborg University and Chalmers University of Technology,}
\centerline{\it S-412 96 G\"oteborg, Sweden.}
\medskip
\centerline{\tt tfemc@fy.chalmers.se}
\vskip1.2cm
\centerline{\twelverm Anders Westerberg}
\vskip.6cm
\centerline{\it DAMTP, University of Cambridge,}
\centerline{\it Silver St., Cambridge CB3 9EW, U.K.}
\medskip
\centerline{\tt A.Westerberg@damtp.cam.ac.uk}
\catcode`\@=\active

\vskip2.5cm

\centerline{\bf Abstract}

{\narrower{\narrower\noindent 
We propose a method for constructing super-brane actions where every
background tensor potential corresponds to a world-volume field strength.
The procedure provides a natural coupling to the background
and automatically displays the ${\rm SL}(2;\bZ)$ symmetry of the 
\II B string theory.
The Dirichlet {\old3}-brane is used as a test ground for these ideas. 
A polynomial
action consistent with non-linear self-duality is presented. 
Invariance of the action under $\k$-symmetry is demonstrated for arbitrary
on-shell type \II B supergravity backgrounds and is shown to require
self-duality.
\smallskip}\smallskip} 

\vfill

\eject

\def\nl{\hfill\break\indent}
\def\nlni{\hfill\break}

\ref\Polchinski{J. Polchinski, {\xit ``Dirichlet-branes and Ramond--Ramond 
	charges''},
	\nl Phys.~Rev.~Lett.~{\xbf 75} ({\xold1995}) {\xold4724} 
	({\xtt hep-th/9510017}).}
\ref\DLP{J. Dai, R.G. Leigh and J. Polchinski,
	{\xit ``New connections between string theories''}, \nl
	Mod.~Phys.~Lett.~{\xbf A4} ({\xold1989}) {\xold2073}.}
\ref\Leigh{R.G. Leigh, {\xit ``Dirac--Born--Infeld action from Dirichlet sigma
	model''}, Mod.~Phys.~Lett.~{\xbf A4} ({\xold1989}) {\xold2767};\nlni
	C.G. Callan, C. Lovelace, C.R. Nappi and S.A. Yost,
	{\xit ``String loop corrections to beta functions''},
	\nl Nucl.~Phys. {\xbf B288} ({\xold1987}) {\xold525}.}
\ref\CvGNW{M.~Cederwall, A.~von~Gussich, B.E.W.~Nilsson and A.~Westerberg,\nl
	{\xit ``The Dirichlet super-three-brane in type IIB supergravity''},
	\nl Nucl. Phys. {\xbf B490} ({\xold 1997}) {\xold 163}
        ({\xtt hep-th/9610148}).}
\ref\CvGNPW{M. Cederwall, A. von Gussich, B.E.W. Nilsson, P. Sundell
        and A. Westerberg, \nl{\xit ``The Dirichlet super-p-branes in
        type \II A and \II B supergravity''}, 
	\nl Nucl. Phys. {\xbf B490} ({\xold 1997}) {\xold 179}
        ({\xtt hep-th/9611159}).}
\ref\agan{M. Aganagic, C. Popescu and J.H. Schwarz, 
	{\xit ``D-brane actions with local kappa symmetry''}, 
	\nl Phys. Lett. {\xbf B393} ({\xold 1997}) {\xold 311}
	({\xtt hep-th/9610249});\nl
        {\xit ``Gauge-invariant and gauge-fixed D-brane actions''},
        Nucl. Phys. {\xbf B495} ({\xold 1997}) {\xold 99}
	({\xtt hep-th/9612080}).}
\ref\bergstown{E. Bergshoeff and P.K. Townsend, {\xit ``Super D-branes''}, 
	Nucl. Phys. {\xbf B490} ({\xold 1997}) {\xold 145} 
	({\xtt hep-th/9611173}).}
\ref\HullTownsend{C.M.~Hull and P.K.~Townsend, 
	{\xit ``Unity of superstring dualities''}, 
	\nl Nucl.~Phys.~{\xbf B438} ({\xold1995}) {\xold109} 
	({\xtt hep-th/9410167}).}
\ref\Schwarz{J.H. Schwarz, {\xit ``An {\xrm SL(2;{\eightbbb Z})} multiplet of 
	type \II B superstrings''}, 
	Phys. Lett. {\xbf B360} ({\xold 1995}) {\xold 13}
	\nl({\xtt hep-th/9508143});
        Erratum: ibid. {\xbf B364} ({\xold 1995}) {\xold 252}.} 
\ref\PKT{P.K. Townsend, {\xit ``Membrane tension and manifest
        \II B S-duality''}, {\xtt hep-th/9705160}.}
\ref\CT{M. Cederwall and P.K. Townsend, {\xit ``The manifestly 
	{\xrm SL(2;{\eightbbb Z})}-covariant superstring''}, 
	\nl{\xtt hep-th/9709002}, to appear in JHEP.}
\ref\Witten{E. Witten, {\xit ``Bound states of strings and p-branes''},
        Nucl. Phys. {\xbf B460} ({\xold 1996})  {\xold 335}
	({\xtt hep-th/9510135}).}
\ref\Tseytlin{A.A.~Tseytlin, {\xit ``Self-duality of Born--Infeld action and 
	Dirichlet 3-brane of Type IIB superstring''},
	\nl Nucl.~Phys.~{\xbf B469} ({\xold1996}) {\xold51}
	({\xtt hep-th/9602064});\nlni
	M.B.~Green and M.~Gutperle, {\xit ``Comments on 3-branes''},
	Phys.~Lett.~\xbf B377 \xrm ({\xold1996}) {\xold28}
	({\xtt hep-th/9602077}).\vfill\eject}
\ref\PSTmfl{ P. Pasti, D. Sorokin and M. Tonin, {\xit ``On Lorentz invariant 
	actions for chiral p-forms''},\nl Phys. Rev. {\xbf D55} ({\xold1997}) 
	{\xold6292} ({\xtt hep-th/9611100});\nlni
	B. McClain, Y.S. Wu and F. Yu, {\xit ``Covariant quantization 
	of chiral bosons and {\xrm OSp(1,1$\ss|$2)} symmetry''},
	\nl Nucl. Phys. {\xbf B343} ({\xold1990}) {\xold689};\nlni
	N. Berkovits, {\xit ``Local actions with electric and magnetic 
	sources''},\nl Phys. Lett. {\xbf B395} ({\xold1997}) {\xold28}
	({\xtt hep-th/9610134});\nlni
	I. Bengtsson, {\xit ``Manifest duality in Born--Infeld theory''},
        {\xtt hep-th/9612174}.}
\ref\Wittenone{E.~Witten, {\xit ``Five-brane effective action in M theory''},
	{\xtt hep-th/9610234}.}
\ref\JHS{J.H. Schwarz, {\xit ``Covariant field equations of chiral N=2 D=10
        supergravity''}, Nucl. Phys. {\xbf B226} ({\xold 1983}) {\xold 269}.}
\ref\HoweWest{P.S.~Howe and P.C.~West, 
	{\xit ``The complete N=2, d=10 supergravity''},
	Nucl.~Phys.~{\xbf B238} ({\xold1984}) {\xold181}.} 
\ref\Khoudeir{D.S. Berman, {\xit ``{\xrm SL(2;{\eightbbb Z})} duality 
	of Born--Infeld theory from self-dual electrodynamics 
	in 6 dimensions''},\nl {\xtt hep-th/9706208}; \nlni
	A. Khoudeir and Y. Parra,
	{\xit ``On duality in the Born--Infeld theory''}, 
	{\xtt hep-th/9708011}.}
\ref\BLT{P.K. Townsend, {\xit ``Worldsheet electromagnetism and the superstring
        tension''}, Phys. Lett. {\xbf 277B} ({\xold 1992}) {\xold 285};\nlni
        E. Bergshoeff, L.A.J. London and P.K. Townsend,
        {\xit ``Space-time scale-invariance and the super-p-brane''},\nl
        Class. Quantum Grav. {\xbf 9} ({\xold 1992}) {\xold 2545} 
	({\xtt hep-th/9206026}).}
\ref\BrinkHowe{E.~Cremmer and S.~Ferrara, 
	{\xit ``Formulation of eleven-dimensional supergravity 
	in superspace''},\nl Phys.~Lett.~{\xbf 91B} ({\xold1980}) {\xold61};
	\nlni L.~Brink and P.~Howe, {\xit ``Eleven-dimensional supergravity 
	on the mass-shell in superspace''},
	\nl Phys.~Lett.~{\xbf 91B} ({\xold1980}) {\xold384}.}
\ref\Candiello{A.~Candiello and K.~Lechner, {\xit ``Duality in supergravity
	theories''},\nl Nucl.~Phys.~{\xbf B412} ({\xold1994}) {\xold479}
	({\xtt hep-th/9309143}).}
\ref\BST{E.~Bergshoeff, E.~Sezgin and P.K.~Townsend, 
	\nl {\xit ``Supermembranes and eleven-dimensional supergravity''},
	Phys.~Lett.~{\xbf B189} ({\xold1987}) {\xold75};
	\nl {\xit ``Properties of the eleven-dimensional supermembrane 
	theory''}, Ann.~Phys. {\xbf 185} ({\xold1988}) {\xold330}.}
\ref\Fivebrane{P.S.~Howe and E.~Sezgin,
	{\xit ``d=11,p=5''}, Phys. Lett. {\xbf B394} ({\xold1997}) {\xold62}
	({\xtt hep-th/9611008});\nlni
	I. Bandos, K. Lechner, A. Nurmagambetov, P. Pasti, D. Sorokin
	and M. Tonin,\nl {\xit ``Covariant action for the super-five-brane 
		of M-theory''},\nl
	Phys. Rev. Lett. {\xbf78} ({\xold1997}) {\xold4332}
		({\xtt hep-th/9701037});\nlni 
	M. Aganagic, J. Park, C. Popescu and J.H. Schwarz,
	{\xit ``World-volume action of the M theory five-brane''},
	\nl Nucl. Phys. {\xbf B496} ({\xold1997}) {\xold191}
		({\xtt hep-th/9701166}).}


\section\Intro{Introduction}Extended supersymmetric objects arise as
soliton configurations in the low-energy effective field theories of
string theory and M-theory. 
Most profoundly, the solitons of type \II\ string theory are realised
as D-branes~[\Polchinski], \ie\ objects on which elementary strings may
end~[\DLP].
Many aspects of the dynamics of these objects, which is the main
subject of this paper, are by now fairly well understood. For instance, 
it was shown early on that the fact that elementary strings may end on
a D-brane necessitates the presence of a self-interacting vector potential on
the D-brane world-volume with its dynamics 
governed by the Dirac--Born--Infeld (DBI) action~[\Leigh]. 
More recently, the world-volume dynamics of 
supersymmetric D-branes has been completely formulated
[\CvGNW-\bergstown]. An important ingredient in the latter construction
is $\k$-symmetry, a fermionic gauge symmetry reflecting the fact that
these configurations are BPS-saturated and break half the space-time
supersymmetries. 

A serious drawback of the picture of refs.~[\CvGNW-\bergstown] for the
case of type \II B branes is that it breaks the SL(2;$\Z$) S-duality
symmetry~[\HullTownsend] of the type \II B theory. 
This shortcoming has more or less been taken for granted, since SL(2;$\Z$) 
is a non-perturbative symmetry and a choice of elementary excitations
is required; indeed, the non-covariance is inherent in the very notion
of D-branes as objects on which elementary open strings may end. 
Nevertheless, recent developments~[\Schwarz,\PKT,\CT] have shown that 
by treating simultaneously the entire SL(2;$\Z$) multiplet of type \II B
superstrings as ``elementary'', a formulation that makes the 
SL(2;$\Z$) symmetry manifest can be achieved.
While such a description should not be expected to be possible in
terms of local field theory for a theory possessing a strong-weak coupling 
symmetry involving the interchange of electric and magnetic charges
(such as, \eg, $N$=4 super Yang--Mills theory), 
the string theory symmetry does not involve any such charge interchange, 
and there seems to be no immediate obstruction to an 
SL(2;$\Z$)-covariant {\it perturbative} formulation.	

The other branes of the type \II B theory also come in SL(2;$\Z$) 
multiplets, and one would like to find a description that respects 
this covariance, 
generalising the superstring action of ref.~[\CT]. The present paper makes
a concrete proposal for how this may be accomplished.
Our main proposal is that any tensor potential in the background
supergravity theory should correspond to a world-volume field strength
containing the background potential schematically as ``$F\is dA\minus C$'',
thus generalising the mechanism required to achieve gauge invariance in 
the traditional formulation.  
If the background theory (type \II B supergravity) is formulated with 
manifest SL(2;$\R$) symmetry, 
the same will then be true for the classical brane action. Quantum effects
on the superstring world-sheet break SL(2;$\R$) to 
SL(2;$\Z$)~[\HullTownsend,\Witten,\CT]. 

The formulation is not limited to type \II B, but should have
its most important application there. In addition
to displaying more symmetry, it should provide clearer information
about branes ending on other branes, in which case the boundary should
couple directly to a tensor potential on the brane. 
It may also simplify the formulation
in other cases (type \II A and {\old11} dimensions), since it in general seems
to yield polynomial actions.

In the present paper, our proposal is tested on the {\old3}-brane of type \II B
string theory, which is a singlet under SL(2;$\Z$). 
This fact has previously been shown to correspond to certain duality 
properties of the DBI action~[\Tseytlin],
and those results are now recast in a completely covariant and  
SL(2;$\Z$)-symmetric form, including the coupling to the background.
The type \II B {\old5}-brane case, which will be even more interesting
since the charges come in an SL(2;$\Z$) doublet, will follow a similar
pattern and will be worked out in a separate paper. 

We should mention that the action we write down does not imply the
non-linear self-duality constraint that is imposed on the world-volume
{\old2}-form field strength. 
There are methods for incorporating self-duality so that it follows from an
action principle [\PSTmfl], but none of these seem to clarify any
of the points we want to make. Indeed, consistency of self-duality with
the Bianchi identities and equations of motion turns out to be a
powerful instrument by itself. If self-duality can be handled in
functional integration, as described by Witten for the (linearised) 
{\old11}-dimensional {\old5}-brane [\Wittenone], 
a formulation like the present one is equally useful.
 
The disposition of the paper is as follows. In section {\old2} we recapitulate
some essentials of type \II B supergravity, and work out the exact structure
of the world-volume fields. section {\old3} deals with the non-linear 
self-duality relevant for the {\old3}-brane, and casts it into a covariant 
form.
In section~{\old4} the {\old3}-brane action is given, and in section {\old5}
$\k$-symmetry is used to derive the same form of the action.
Finally, section {\old6} contains a discussion of our results together
with some conjectures on the further application of our method.
Some notes on notation are found in an appendix.

\section\Fields{World-volume and background fields}We advocate the proposal
that coupling of branes to background tensor potentials of any rank should be 
achieved solely through the formation of world-volume field strengths
of the type ``$F\is dA\minus C$'', $C$ being (the pullback of) a 
background potential
with field strength ``$H\is dC$'' and $A$ a world-volume potential. 
The quotation marks are meant to indicate that these relations are to
be seen as schematic; the actual background field theories, supergravities
in {\old10} or {\old11} dimensions, contain ``modified'' fields strengths, 
a fact which
will imply corresponding deviations from the schematic expressions
above also on the world-volume. We will concretise all this in a little
while for the type \II B theory; we expect the type \II A and 
{\old11}-dimensional cases to behave similarly. 

Before investigating the structure of the world-volume fields, some words
need to be said about the background. In the present paper, we will focus
on the case of the {\old3}-brane in type \II B string theory and the relevant
background fields will be those of type \II B supergravity [\JHS,\HoweWest].

The type \II B branes are embedded in chiral complex $(10|32)$ superspace 
with coordinates $Z^M\is(X^m,\theta^\mu,\bar\theta^{\bar\mu})$. Local Lorentz
indices are $A\is(a,\a,\bar\a)$. To avoid confusion, note that although 
the indices $\a$ and $\bar\a$ denote components of complex spinors, we
may still use real (Majorana) $\g$-matrices, which makes them behave 
identically under local Lorentz transformations. In some cases, such as the 
supergravity constraints of section {\old5}, objects with seemingly different
index structures may thus be identified.

The bosonic sector of type \II B supergravity consists, apart from gravity, 
of scalar fields in the coset ${\rm SL}(2;\R)/{\rm SO}(2)$, 
an SL$(2;\R)$ doublet of {\old2}-form potentials
$C_r$ and an SL$(2;\R)$ singlet {\old4}-form potential $C$ with self-dual
field strength. The formulation of this theory is well known, and we will only
briefly state some relevant points in order to establish our notation.

The scalars are represented as a complex SL(2;$\R$) 
doublet $\U^r$ constrained by
${i\/2}\e_{rs}\U^r\bar\U^s\is1$. The coset is obtained by gauging the
U(1) acting as $\U^r\!\rightarrow\!e^{i\vartheta}\U^r$.
We normalise the U(1) charge $e$ to 1 for the scalars $\U^r$.
The physical dilaton and axion fields $\phi$ and $\chi$ are identified as 
the projective invariant $\U^2\!/\U^1\is\chi\plus ie^{-\phi}$, but we will
stick to the SL$(2;\R)$-covariant notation.
If the above action of SL$(2;\R)$ on the scalars is considered the left action,
there is also a right action, of which the U(1) is a subgroup.
The left-invariant Maurer--Cartan forms and equations are
$$
\hbox{\vtop{
\halign{$#$\hfill\qquad&$#$\hfill\cr
Q=\half\e_{rs}d\U^r\bar\U^s\komma	&dQ-iP\w\bar P=0\komma\ccr
P=\half\e_{rs}d\U^r\U^s\komma		&dP-2iP\w Q\equiv DP=0\komma\cr}
}}\eqn
$$
where the covariant derivative $D$ (acting from the right) includes
the real U(1) connection $Q$ as $D=d-ieQ$, $e$ denoting the U(1) charge.

The {\old3}-form field strengths are formed as $H_r=dC_r$. 
The scalars act as a bridge between objects that are SL$(2;\R)$ doublets
and U(1) singlets, such as $H_r$, and the corresponding objects
which are SL$(2;\R)$ singlets but carry U(1) charge 1, such as
$\H\equiv H_r\U^r$. In this function they are necessary for the existence
of SL$(2;\R)$-invariant kinetic terms like $\H\dot\bar\H$ (for the $\theta\is0$
components).
We are using roman versus script letters to make
the same distinction also for other fields.
The complex field strength satisfies the Bianchi identity
$$
dH_r\U^r\equiv D\H+i\bar\H\w P=0\punkt\eqn
$$
The {\old5}-form field strength and Bianchi identity are
$$
H=dC+\Im(\C\w\bar\H)\komma\qquad dH-i\H\w\bar\H=0\punkt\Eqn\FiveformBI
$$
where the ``modifying'' terms come from consistency of self-duality
with the presence of the supergravity ``Chern--Simons'' term
$\propto i\int\H\w\bar\H\w C$. A corresponding statement of course applies
for the {\xold7}-form, whose $\theta\is0$ component is dual to the 
{\old3}-form.
Analogous consistency conditions imposed on world-volume fields will
be central for the continued discussion.

With these preliminaries at hand, we are ready to treat world-volume fields
defined by modifications of the typical relation ``$F\is dA\minus C$''.
The guiding principle is gauge invariance, both with respect to background
and world-volume gauge transformations. Up to trivial rescalings, one is 
uniquely led to the {\old2}-form and {\old4}-form field strengths
$$
\F=dA_r\U^r-\C\komma\qquad F=dA-C+\Im(\A\w\bar\H)\komma\eqn
$$
with the Bianchi identities
$$
D\F+i\bar\F\w P=-\H\komma\qquad dF=-H-\Im(\F\w\bar\H)\punkt\eqn
$$
It is straightforward to verify that all field strengths are invariant 
under the gauge transformations
$$
\eqalign{
&\d C_r=dL_r			\komma\cr
&\d C=\Im(\L\w\bar\H)+dL	\komma\cccr
&\d A_r=dl_r+L_r		\komma\cr
&\d A=\Im(\ell\bar\H)+dl+L	\komma\cr
}\Eqn\GaugeTransformations
$$
where $\L\!\equiv\!L_r\U^r$ and $L$ are background {\old1}- and {\old3}-form 
gauge parameters,
while the world-volume parameters 
$\ell\!\equiv\!l_r\U^r$ and $l$ are {\old0}- and {\old2}-forms.
These definitions are independent of the dimension of the world-volume.
For the case of a {\old3}-brane these fields are sufficient; for 
higher-dimensional $p$-branes the procedure will and can be continued
up to the field strength of rank $p\plus2$.

The number of world-volume fields and their transformation properties
are thus completely determined by the structure of the background 
supergravity theory, in this case type \II B. We know, on the other hand,
\eg\ by comparison to the field content of D-branes, that this in general
gives too many bosonic degrees of freedom. Apart from transverse
oscillations, type \II\ branes contain one vector potential. The 
{\old3}-brane, 
following our proposal, carries an SL$(2;\R)$ doublet of vector
potentials and a {\old3}-form without local degrees of freedom, while a 
{\old5}-brane 
has the vector doublet, the {\old3}-form singlet and a doublet of {\old5}-form
potentials without local degrees of freedom. Thus, there must be a mechanism
by which the number of physical fields is reduced to the correct one.
This mechanism is a kind of duality relation, which will be investigated for
the {\old3}-brane in the next section. We would like to remark that even if
the proposed set of fields containing background couplings may not be seen
as necessary, manifest SL$(2)$ symmetry requires at least the doublet
of vector potentials, which already calls for some relation between
the field strengths. The exception is the case $p\is1$ treated in 
refs.~[\PKT,\CT], 
where the vector potentials do not carry any local degrees of freedom.

\section\Duality{Self-duality}The duality properties of the 
DBI action for the {\old3}-brane are well understood; 
the {\old3}-brane
is invariant under a simultaneous duality transformation of the 
world-volume vector field 
and a transformation of the background fields under the non-trivial 
element in the $\Z_2$ subgroup of SL$(2;\Z)$ interchanging NS-NS and 
RR charges [\Tseytlin]. 
The scalar fields have also been included in this transformation [\Khoudeir], 
but not in a manifestly SL$(2;\Z)$-covariant way. Neither have the
duality properties been examined in the supersymmetric case, though it is
obvious that they must hold. We will cast the known duality transformations
into a covariant form, to be used later for a manifestly SL$(2)$-covariant
formulation of the supersymmetric {\old3}-brane.

The {\xold3}-brane DBI action 
$$
S_{\rm DBI} = - \int\!d^4\xi\,\sqrt{-\det(g+F)}\,\komma\Eqn\DBIAction
$$
contains one real {\old2}-form field strength $F$. 
The dual field strength $G$, 
defined by $*G=\pm2{\d S_{\hbox{\fiverm DBI}}\/\d F}$,
is related to $F$ by the non-linear duality transformation
$$
G=\pm N^{-1}(F)\left({*}F+\half(F\dot{*}F)F\right)\komma\Eqn\RealDuality
$$
with inverse
$$
F=\mp N^{-1}(G)\left({*}G+\half(G\dot{*}G)G\right)\komma\Eqn\RealDualityInverse
$$
where $N(F)=(1+F\dot F-\frac4(F\dot{*}F)^2)^{1/2}$ 
(note that $\sqrt{-g}N(F)=-\L_{\rm DBI}$).

We would like to associate these real fields
with the real components of the field strength $\F$ of the previous 
section. The first critical step to be taken is to check whether 
(\RealDuality) respects the U(1) symmetry. The duality transformation
is not manifestly U(1) symmetric, but it is quite straightforward
to show that it holds unchanged also for the real components of
$e^{i\vartheta}(F\plus iG)$. Disregarding the physical reasons for identifying
$\F\is F\plus iG$, this is quite astonishing, since (\RealDuality) contains
an infinite number of terms with different U(1) charges. The next task
is to find the manifestly U(1)-symmetric formulation of the duality.
One piece of information is found in the observation that
$G\dot{*}G=-F\dot{*}F$, a relation which takes the U(1)-covariant form 
$$
\F\dot{*}\bar\F=0\punkt\Eqn\ZeroBilinear
$$ 
We therefore make an ansatz for self-duality 
consistent with this identity:
$$
i{*}\F=\a\F+\b\left((\F\dot\F)\bar\F+(\F\dot{*}\F){*}\bar\F\right)\komma
\Eqn\Zdual
$$
where $\a$ and $\b$ are uncharged scalar functions of $\F$. 
This ansatz must be restricted
in order to be consistent (eq.~(\Zdual) encodes both the duality transformation
(\RealDuality) and its inverse (\RealDualityInverse)). In order to shorten
the expressions, we introduce the quadratic invariants 
$r\!\equiv\!\F\dot\bar\F$, $p\!\equiv\!\F\dot\F$, $q\!\equiv\!\F\dot{*}\F$ 
at U(1) charges 0, 2 and 2, respectively. Consistency of the non-linear 
self-duality demands that the relations
$$
\eqalign{
	&q=-i(\a+\b r)p\quad\left(=-{i\/\a-\b r}p\,\right)\komma\cr
	&r=i\b(p\bar q-\bar pq)\quad
		\left(=-{\b\/\a}(p\bar p+q\bar q)\,\right)\komma\cccr
	&\a^2-\b^2r^2=1\cr
}\Eqn\pqrabrelations
$$
be satisfied. This leaves only one independent complex invariant,  
matching in number the two scalars $F\dot F$ and $F\dot{*}F$. 
The remaining freedom resides
in the choice of the function $\b$, which will be determined from the actual
duality (\RealDuality). Since we have U(1) covariance, we may choose
any equation following from (\RealDuality) but not implied by (\ZeroBilinear)
that we wish. Even if it consists of terms with different charge $e$, we
know that it will be fulfilled separately at each value of $e$. 
We chose to use the symmetric equation 
$N(F)N(G)=1+\frac{16}(\Re q)^2$. We also checked explicitly that any
of the charge components of this equation implies
$$
\a=\pm {\,\,\,\,1+{r\/8}\/\sqrt{1+{r\/4}\,}}\komma\qquad
\b=\mp{1\/8\,\sqrt{1+{r\/4}\,}}\punkt\Eqn\absolutions
$$
The duality may be simplified further; one useful observation, that we will
employ in the proof of $\k$-symmetry as well, is that 
when the self-duality constraint is satisfied, each of the matrices
${\F_i}^j$, $\bar\F_i\hskip.6pt{}^j$, ${*}{\F_i}^j$ and 
${*}\bar\F_i\hskip.6pt{}^j$ is 
expressible as a sum with scalar coefficients of odd powers of only one
real matrix ${F_i}^j$, and, consequently, they all commute. Other relations
that follow are (matrix multiplication) $\F{*}\F\is{1\/4}\tr(\F{*}\F)\id$ and
$\F{*}\bar\F\is-{*}\F\bar\F$. The final form of the self-duality relation is
$$
{*}\F={\mp i\/\sqrt{\,1+{1\/4}\F\dot\bar\F\,}}
\left(\F-\frac4(\F\dot{*}\F){*}\bar\F\right)\punkt
\Eqn\Dualityrelation
$$ 
In the following section we will formulate an action that is consistent
with this relation.

\section\Action{The 3-brane action}There is a number of 
qualifications we require of the action, 
the first one being, of course, that it be
equivalent to the DBI action. This may only be achieved if the action 
is compatible with 
the self-duality relation (\Dualityrelation), which is what guides us
in the first place. We also want manifest SL(2)-covariance, which is
achieved by using only the fields of section \Fields.
Finally, we need $\k$-symmetry to get the correct number of fermions
on the world-volume.

In the spirit of earlier formulations of strings and membranes with
$p$-form world-volume potentials [\BLT,\PKT,\CT], 
we make an ansatz for the action as
$$
S=\int\!d^4\xi\,\l\sqrt{-g}\left\{1+\Phi(\F,\bar\F)+cF\dot F\right\}
	\komma\Eqn\ActionAnsatz
$$
where $\l$ is a Lagrange multiplier field and $c$ a numerical constant.
In order to judge the consistency of self-duality, we need to compare
Bianchi identities and equations of motion for the field strength $\F$.
To this end, define 
$\K\is{\*\Phi\/\lower1.5pt\hbox{$\ss\*\bar\F$}}$. 
Noticing that the last term in the action (\ActionAnsatz) may
be recast in the alternative forms
$-c\int\! d^4\xi\l\sqrt{-g}({*}F)^2\is -c\int\!\l F\w{*}F$, 
the equation of motion for $\l$ is readily found to be
$$
{*}F=\pm{1\/\sqrt{c}}\sqrt{1+\Phi}\komma\Eqn\FSolution
$$
while the one for $A$ reads $d(\l{*}F)\is0$. Furthermore, taking into
account the fact that $F$ contains $\A$, 
the equations of motion for $A_r$ may be written
$$
D(\l{*}\K)-i\l{*}\bar\K\w P=ic\l({*}F)\H\komma\Eqn\PreliminaryDual
$$
which should be compared to the Bianchi identity $D\F\plus i\bar\F\w\,P\is-\H$.
The only duality relation consistent with the equations of motion 
following from the action (\ActionAnsatz), in the sense that the 
background field $\H$ cancels, is 
$$
{*}\K=-ic\F({*}F)\punkt\Eqn\ImplicitDuality
$$ 
Note that the sign of the term containing the Maurer--Cartan form $P$ changes
under multiplication of the field with $i$. This is the only check we
need to perform in order to see that duality works for arbitrary
background dilaton and axion fields.
Matching eq.~(\ImplicitDuality) 
with the results of the previous section, we arrive at $c\is{1\/4}$ 
(the actual number being, of course, a consequence of the choice 
of normalisation
in (\FiveformBI)) and
$$
S=\int\!d^4\xi\,\l\sqrt{-g}\Bigl\{1+\half \F\dot\bar \F
   	-\frac{16}(\F\dot{*}\F)(\bar \F\dot{*}\bar \F)
	+\frac4F\dot F\Bigr\}\punkt\Eqn\FinalAction
$$
The duality now takes the polynomial form
$$
\ihalf({*}F){*}\F=\F-\frac4(\F\dot{*}\F){*}\bar\F\punkt\Eqn\PolynomialDuality
$$
Note that once the duality is satisfied, the quartic term 
in (\FinalAction) reduces to
$-{1\/4}\F\dot\bar \F$, so that the square root in (\FSolution) becomes
identical to the one in (\Dualityrelation) (this relation should of course not
be inserted into the action).
It is, in fact, not necessary to determine the unknown ingredients in
(\ActionAnsatz) by comparison with earlier results. One may as well
go directly to the following section and get the form (\FinalAction)
directly from the requirement of $\k$-symmetry. 

Let us end the present section with a comment on the fact that there is no 
Wess--Zumino term in the action. Considering the actual
form of $F$ and its equation of motion, a partial integration is sufficient
to see that the $F^2$ term plays, in fact, the same r\^ole as
the Wess--Zumino term does in the ``old'' D{\old3}-brane action 
(although there is
a difference in that the present term includes both {\old2}-form 
field strengths). 
This statement is true in general
for the $(p\plus1)$-form field strengths.
The sign of the duality (\PolynomialDuality) depends on the sign of ${*}F$
in eq.~(\FSolution). Therefore the present action describes two sectors
of solutions, where the {\old2}-form is self-dual or anti-self-dual, 
\ie\ {\old3}-branes
and anti-{\old3}-branes. In previous descriptions the two sectors differ
by the sign of the Wess-Zumino term, or, equivalently, of the ``RR'' charge
(the notion is really an artefact of the SL$(2;\Z)$-non-covariant
choice of elementary string excitations; for the {\xold3}-brane this
charge is SL$(2;\Z)$-invariant).

\section\Symmetry{{\twelvemath\char'24}-symmetry}$\k$-symmetry is the 
invariance under the transformations induced by
the local fermionic translations
$\d_\k Z^M\is\k^\a{E_\a}^M\plus\,\bar\k^{\bar\a}{E_{\bar\a}}^M$. 
In particular, for a background form pulled back to the
world-volume the induced transformation reads
$\d_\k\Omega\is\L_\k\Omega\is(i_\k d+di_\k)\Omega$.
Transformations of world-volume
fields must be specified separately, with the purpose of making the
transformations gauge-covariant. To this end, the potentials of
section \Fields\ are taken to transform as
$$
\eqalign{
&\d_\k A_r=i_\k C_r\komma\cr
&\d_\k A=i_\k C-\Im(\A\w i_\k\bar\H)	\komma\cr
}\Eqn\PotentialKappa
$$
which results in the following transformations of the world-volume
field strengths:
$$
\eqalign{
&\d_\k\F=-i\bar\F\w i_\k P-i_\k\H	\komma\cr
&\d_\k F=-i_\k H-\Im(\F\w i_\k\bar\H)\punkt\cr
}\Eqn\FieldKappa
$$
One also has to consider the transformation of the induced metric,
$$
\d_\k g_{ij}=2{E_{(i}}^a{E_{j)}}^B\k^\a\,{T_{B\a}}^b\eta_{ab}+\cc\eqn
$$
The gauge parameter $\k$ is subject to some restriction that makes it 
gauge away only half the fermions; an essential part of establishing 
$\k$-symmetry amounts to finding the half-rank matrix annihilating $\k$.

When the above transformation rules are applied in the variation of an action,
one needs explicit expressions for the background field components involved. 
The latter are only those carrying at least one spinor index, and as
such they are subject to constraints. A convenient set of constraints,
which is consistent with the Bianchi identities of section {\old2} as
well as with those for the torsion and implies the equations of motion
for the background fields, is at dimension 0:
$$
\eqalign{
&{T_{\a\bar\b}}^a=(\g^a)_{\a\b}\komma\cr
&\H_{a\a\b}=2(\g_a)_{\a\b}\komma\cr
&H_{abc\a\bar\b}=2i(\g_{abc})_{\a\b}\komma\cr
}\Eqn\DimZeroConstraints
$$
and at dimension 1/2:
$$
\eqalign{
	&H_{abcd\a}=0\komma\cr
        &\H_{ab\bar\a}=-i(\g_{ab}P)_\a\komma\cr
        &{T_{\bar\a\bar\b}}^\g=i{\d_{(\a}}^\g P_{\b)}
                -\ihalf(\g_a)_{\a\b}(\g^aP)^\g\komma\cr
        &P_{\bar\a}=0\, ,\qquad Q_\a=0\semikolon\cr}\Eqn\DimHalfConstraints
$$
the constraints at dimension 1 and above do not enter in the $\k$-variation. 
Components of the appropriate dimensions not appearing in (\DimZeroConstraints)
or (\DimHalfConstraints) vanish. Also, we do not separately give
components that follow by complex conjugation. 

At this point, it is very convenient to take advantage of the fact
that we have a duality
relation; since the variation of the action, irrespective of the exact
form of the latter, will
contain the fields $\K$ of section \Action, we may use self-duality in
the form (\ImplicitDuality) to obtain the exact variation whatever the 
function $\Phi$ may be. The variation of the Lagrange multiplier is
as usual unimportant, so we content ourselves with varying the constraint
$\Psi\equiv1+\Phi+{1\/4}F\dot F\approx0$.
The result is at dimension 0 and 1/2, respectively,
$$
\eqalign{
&(\d_\k\Psi)_{(0)}=({*}F)\bar E_i\left\{\left[\,\half({*}F)\g^i
	-\hbox{${\hbox{\eightmath i}\/6\sqrt{-g}}$}\e^{ijkl}\g_{jkl}
	+\ihalf({*}\F\bar\F)^{ij}\g_j\right]\k
	+i{*}\F^{ij}\g_j\bar\k\,\right\}+\cc\komma\cccr
&(\d_\k\Psi)_{(1/2)}=\frac8(*F)*\F^{ij}\bar P_{\a}
\left[\F_{ij}\bar\k-2(\g_{ij}\k)\right]^{\a}+\cc\cr
}\Eqn\PsiVariation
$$
We must then find a matrix $M$ of half maximal rank such that this variation 
vanishes when ``$\k\is M\zeta$'' (real matrix acting on real components)
for an unconstrained spinor $\zeta$. Indeed, if we let
$$
\k=\left(\hbox{${\hbox{\eightmath i}\/24\sqrt{-g}}$}\e^{ijkl}\g_{ijkl}
+\half{*}F\right)\zeta
	-\ifrac4{*}\F^{ij}\g_{ij}\bar\zeta\komma\Eqn\HalfRank
$$
we find that both parts of (\PsiVariation) vanish, provided that 
$1+{1\/4}\F\dot\bar\F+{1\/4}F\dot F=0$ and the duality relation
(\PolynomialDuality) holds. It is not difficult to see that eq.~(\HalfRank)
can be written as a projection $\k\is P_+\k$, where the projectors $P_\pm$
are given by
$$
({*}F)P_\pm\psi={*}\left(\half F\psi\mp\ihalf\F\w\g_{(2)}\bar\psi
	\pm i\g_{(4)}\psi\right)\punkt\Eqn\PolynomialDuality
$$
The invariance under $\k$-symmetry thus demands the same action and
the same self-duality relation as were derived earlier using 
the DBI action as input. 

\section\Discussion{Discussion}We have demonstrated how the type \II B
string theory super-{\old3}-brane, by the introduction of a full 
set of world-volume
fields corresponding to the background ones, 
can be given a formulation where the SL(2;$\Z$) symmetry becomes
manifest. The construction involved a certain non-linear self-duality
relation needed to reduce the number of physical degrees of freedom 
to the correct one, and also demanded by supersymmetry.
We take this as a strong indication that the idea of introducing
the higher world-volume tensors is indeed a good one. The final test
will of course
be the multiplet of {\old5}-branes, but it is quite clear from the present
investigation how it will work. 
In this case we start from an ansatz of the form 
$$
S=\int\!d^6\xi\,\l\sqrt{-g}\left\{1+\Phi(\F,\bar\F,F)+c\,\G\dot\bar\G\right\}
\komma
$$
where $\G\is dA_{(5)r}\U^r\minus\C_{(6)}\plus{1\/3}A\w\H\minus{1\/6}\A\w H$
is a non-dynamical complex {\old6}-form.
Again, there will be implicit restrictions analogous to (\ImplicitDuality) on
possible duality relations. The exact form of these restrictions is
easy to derive; they imply a duality between the (real) {\old4}-form
and the (complex) {\old2}-form, where one uses $\G$ instead of $\e$
for dualising,
$$
\eqalign{
	&{*}K=-\genfrac{2\eightmath c}3\Re(\F{*}\bar\G)\komma\cr
	&{*}\K=\genfrac{\eightmath c}6F{*}\G
	+\genfrac{\eightmath ic}{12}(\F\w\bar\F{*}\G+\F\w\F{*}\bar\G)\cr
	&\phantom{{*}\K}=\genfrac{\eightmath c}6F{*}\G
		-\ifrac4\F\w{*}K\cr
}\Eqn\FiveDuality
$$
(here, $K={\*\Phi\/\*F}$),
thus resolving the question of how to obtain the correct number
of fields. There are at least three consistency checks involved in the
procedure: the first
one is the internal consistency of the explicit duality ansatz 
analogous to (\Zdual),
the second one its integrability, \ie\ the possibility of deriving it from an
action in the restricted sense described above, and the
third that $\k$-symmetry, which given (\FiveDuality) is calculable without
knowledge of $\Phi$, requires the same duality. 
 
We have in this paper dealt exclusively with type \II B branes, motivated
by the search for SL(2)-invariant formulations. It would be interesting
to see if the method applies also in the cases of type \II A and 
{\old11} dimensions.
This might be done directly in {\old11} dimensions using the superspace
formulation of {\old11}-dimensional 
supergravity [\BrinkHowe,\Candiello] as a background.
The membrane [\BST] only involves a maximum-rank form, 
and is treated in ref.~[\BLT].
It is conceivable that the present method would provide a polynomial 
$\k$-symmetric action consistent with a non-linear self-duality for
the {\old11}-dimensional {\old5}-brane [\Fivebrane].

\vskip\baselineskip\noindent
\hbox{\twelvecp Acknowledgements\hfill}
\vskip.3\baselineskip\noindent
The work of A.W. was supported by the European Commission TMR
Programme under grant ERBFMBI-CT97-2021. M.C. was supported by the
Swedish Natural Science Research Council.

\appendix{Some notation}We use the standard conventions for superspace
forms with an $n$-form expanded as
$$
\Omega={1\/n!}dZ^{M_n}\ldots dZ^{M_1}\Omega_{M_1\ldots M_n}
	={1\/n!}E^{A_n}\ldots E^{A_1}\Omega_{A_1\ldots A_n}\komma\aeqn
$$
and the world-volume forms defined analogously. No separate notation
is used for pullbacks to the world-volume.
The exterior derivative acts from the right as
$$
d\Omega=\Omega\w d\hskip-6.5pt\raise7pt\hbox{$\leftarrow$}\komma\aeqn
$$
and thus obeys 
$$
d(\Omega_{(m)}\w\tilde\Omega_{(n)})=\Omega_{(m)}\w
d\tilde\Omega_{(n)}+ (-1)^n d\Omega_{(m)}\w\tilde\Omega_{(n)}\punkt\aeqn
$$
The interior product $i_V$ by a supervector field acts similarly.
Scalar products of tensors are defined as
$$
A\dot B\equiv{1\/n!}A_{i_1\ldots i_n}B^{i_1\ldots i_n}\komma\aeqn
$$
and Hodge duality as
$$
{*}\Omega^{i_1\ldots i_{D-n}}
\equiv{1\/n!\sqrt{-g}}\e^{i_1\ldots i_n}\Omega_{i_{D-n+1}\ldots i_D}
	\komma\aeqn
$$
with $\e^{01\ldots(D-1)}=1$.

Unlike most authors, we have chosen a convention where complex conjugation does
not reverse the order of fermions, for which reason the dimension-0
component of the torsion in (\DimZeroConstraints) 
does not contain a factor $i$. We find that this
convention simplifies reality properties in complex superspace.

\refout
\end